\documentclass{article}
\usepackage[utf8]{inputenc}
\usepackage{graphicx}
\usepackage{amsmath}
\usepackage{float}
\usepackage{url} 
\usepackage{tabularx}
\usepackage{multicol}
\usepackage[margin=1in]{geometry}
\usepackage{stfloats}
\usepackage[T1]{fontenc}
\usepackage[colorlinks=true,
            linkcolor=blue,      
            citecolor=green,     
            urlcolor=magenta]    
            {hyperref}

\title{Analysis and Resilience of the U.S. Flight Network}
\author{
Shreejan Pandey\thanks{Department of Computer Science, Southern Illinois University Edwardsville. \texttt{shrepan@siue.edu}. These authors contributed equally.}
\and
Sushrit Kafle\thanks{Department of Computer Science, Southern Illinois University Edwardsville. \texttt{sukafle@siue.edu}. These authors contributed equally.}
}

\usepackage{etoolbox}

\begin{document}

\maketitle

\begin{abstract}
Air travel is one of the most widely used transportation services in the United States. This paper analyzes the  U.S. Flight Network (USFN) using complex network theory by exploring how the network's topology contributes to its efficiency and vulnerability. This is done by examining the structural properties, degree distributions, and community structures in the network. USFN was observed to follow power-law distribution and falls under the anomalous regime, suggesting that the network is hub dominant. Compared to null networks, USFN has a higher clustering coefficient and modularity. Various percolation test revealed that USFN is vulnerable to targeted attacks and is susceptible to complete cascading failure if one of the major hubs fails. The overall results suggest that while the USFN is designed for efficiency, it is highly vulnerable to disruptions. Protecting key hub airports is important to make the network more robust and prevent large-scale failures.

\end{abstract}
\begin{multicols}{2}
\section{Introduction}
Air transportation plays a vital role in connectivity by providing rapid movement of goods and people over vast areas. As one of the most used transportation system in the US, it is vital to properly analyze the U.S. Flight Network (USFN). Applying complex network theory and methods for analysis using centralities, we can determine critical hubs and simulate the condition where these nodes are removed. Simulating both random and target attacks to this network can thus provide vital information on networks percolation and resilience analysis. 
Building on the foundation of the research, this paper also provides insight into future works.

\section{Data and Network Construction}
\subsection{Dataset}

The Global Air Transportation Network dataset\footnote{\href{https://www.kaggle.com/datasets/thedevastator/global-air-transportation-network-mapping-the-wo?select=airports.csv}{Global Air Transportation Network – Kaggle}} curated by Tyler Woebkenberg\footnote{\href{https://data.world/tylerudite}{Tyler Woebkenberg's - Data World}} was used for this work.  The dataset includes more than 10,000 data points representing world’s air transportation infrastructure.  The datasets is divided into four main files:

\begin{itemize}
    \item \textbf{\texttt{airports.csv}}: Contains metadata of 7,658 airports worldwide, including name, country, IATA and ICAO codes, geographic coordinates (latitude and longitude), altitude, timezone, and airport type.
    
    \item \textbf{\texttt{routes.csv}}: Describes the flight connections between airports, including the airline operating the route, source and destination IATA codes, number of stops, codeshare information, and aircraft type.
    
    \item \textbf{\texttt{airlines.csv}}: Contains lists of airline operators along with their respective IATA/ICAO codes, country of origin, call signs, and operational status.
    
    \item \textbf{\texttt{airplanes.csv}}: Includes aircraft name with IATA and ICAO codes,.
\end{itemize}

\subsection{Data Filtering}
The initial dataset contains global air transportation data. In order to focus on the U.S flight network only, the dataset went through several filtration steps. Firstly, the \texttt{airports.csv} file was filtered out using the \texttt{Country} column to only include airports located in the United States. Then the \texttt{routes.csv} file was used to retain only those routes where both the source and destination airports were included in the list of U.S. airports filtered previously.  These step ensured that only domestic connections were included in the final network. Finally, in order to remove directionality of the network, the number of occurrences of each route was counted for a given source and destination, and used as the edge weight, hence making the network undirected. 

\subsection{Network Construction}
For the network construction, each airport, denoted by its IATA airport code, was added as a node. Routes between airports were considered as an edge, which connected the two nodes. The number of times each route appeared in the dataset was assigned as the weight for the edge. The final result is a weighted, undirected U.S. Flight Network (USFN), where each node represents an airport and each edge represents a flight connection. The graph was exported as a \texttt{.gml} file for further analysis using Gephi and NetworkX.

\section{USFN Network Property}

\subsection{Nodes and Edges}
\begin{itemize}
  \item In a network, \textbf{node} is an individual entity or object.\\
        In USFN, it represent an \textbf{airport}.

  \item \textbf{Connection} between nodes in a network is called \textbf{edge}.\\
        In USFN edge between two nodes represent a \textbf{flight} between two airport represented by that two node.

  \item \textbf{Weighted edge} refers to an edge with a numeric value associated with it.\\
        USFN consist of weighted edge where \textbf{weight} determine the \textbf{number of flights} between two airport.
\end{itemize}

\subsection{Average Shortest Path Length}
The average number of steps along the shortest path taken by all possible pairs of network nodes. 

Mathematically, 
Average shortest path length \cite{HOSSAIN20171} 
\begin{equation}
L = \frac{1}{n(n - 1)} \sum_{i \ne j} d_{ij}
\label{eq:avg_shortest_path}
\end{equation}
\noindent
where:
\begin{itemize}
  \item $n$ is the total number of nodes in the network,
  \item $d_{ij}$ is the shortest path distance (number of edges) between node $i$ and node $j$,
\end{itemize}
\noindent
The average shortest path length for USFN(L) = 3.2. 

\subsection{Network Diameter}
The longest shortest path between any two nodes in the network. 
\noindent
For USFN the network diameter is 7. This means at max there is at most 7 stops we need to take to travel between two airport. 
\subsection{Degree Distribution}
Degree distribution helps us understand how edge are distributed across nodes. It help us determine critical nodes know as hubs in a network. 
\begin{figure}[H]
    \centering
    \includegraphics[width=\linewidth]{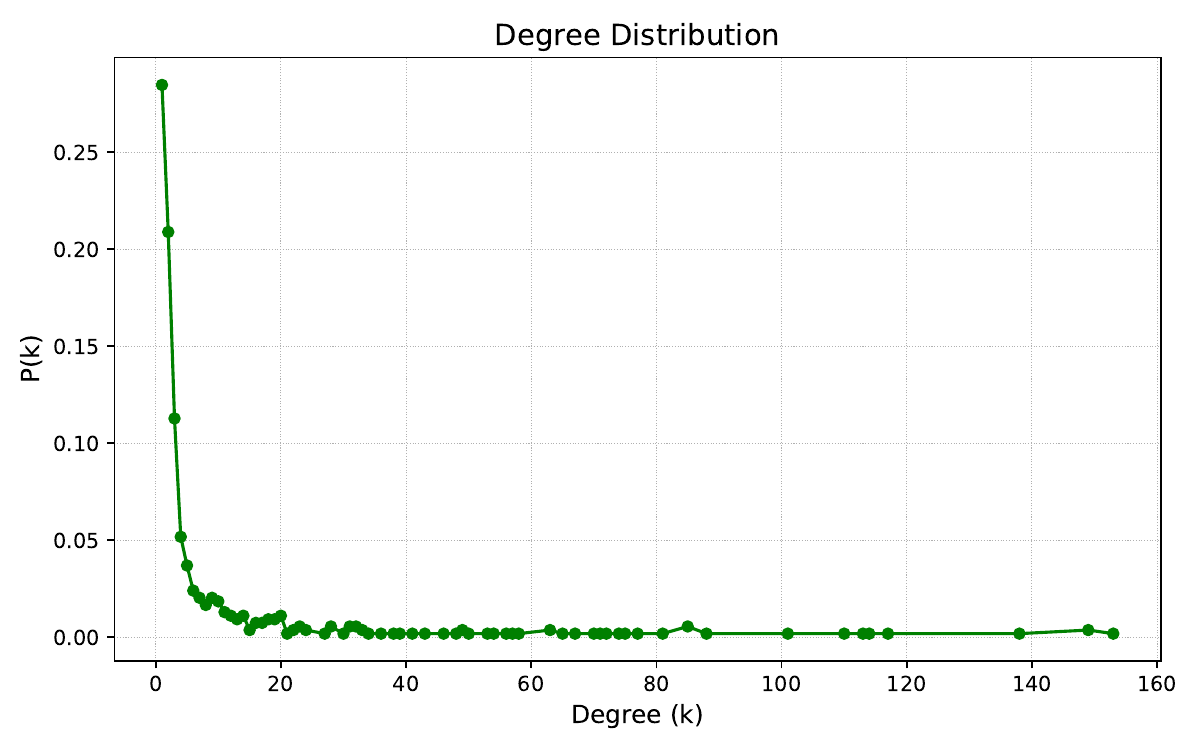} 
    \caption{Degree distribution of the US Flight Network.}
    \label{fig:degree_distribution}
\end{figure}
\noindent
\textbf{Average degree in USFN is 10 and the highest degree in USFN is 153 for node ATL}
\subsection{Betweenness Centrality}

Betweenness centrality for a given node is the amount of time the node is present on the shortest path between any other two nodes.\cite{CURADO2022126560}

Mathematically, given G = (V, E, W), an undirected graph, where V is the set of nodes, E is set of edges and W is the weight of the node. 

Betweenness centrality is given as: \begin{equation}
C_t = \sum_{i \ne t} \sum_{\substack{j \ne i \\ j \ne t}} \frac{\sigma_{itj}}{\sigma_{ij}}.
\end{equation}
where: 
\begin{itemize}
  \item $\sigma_{ij}$ is the total number of shortest paths from node $i$ to node $j$,
  \item $\sigma_{itj}$ is the number of those shortest paths that pass through node $t$.
\end{itemize}

\begin{figure}[H]
    \centering
    \includegraphics[width=0.9\linewidth]{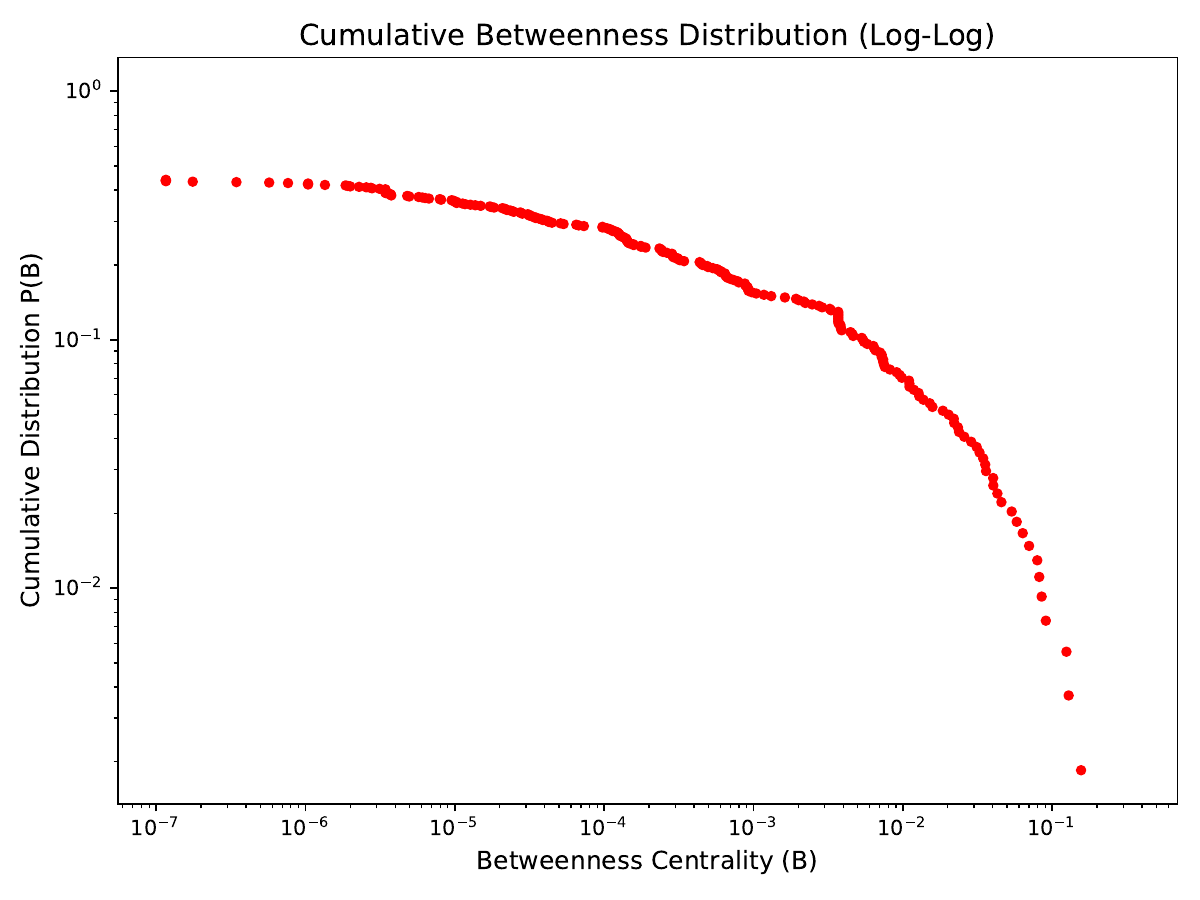}
    \caption{Betweenness centrality of nodes (airports) in the U.S. flight network.}
    \label{fig:betweenness}
\end{figure}

\noindent
\textbf{Ted Stevens Anchorage International Airport(ANC) has the highest betweenness centrality.}
\section{Network Analysis}
\subsection{Power-Law Fitting of the Degree Distribution}
Using the network's degree distribution, we tried to test the scale-free nature of the USFN. A network is scale-free(SF) if the degree distribution follows a power law \cite{Newman_2003}. The power law is represented as: \( P(k) \sim k^{-\gamma} \), where \( P(k) \) is the probability of a node being connected to other k nodes and \( \gamma \) is its degree exponent.
\begin{figure}[H]
    \centering
    \includegraphics[width=0.9\linewidth]{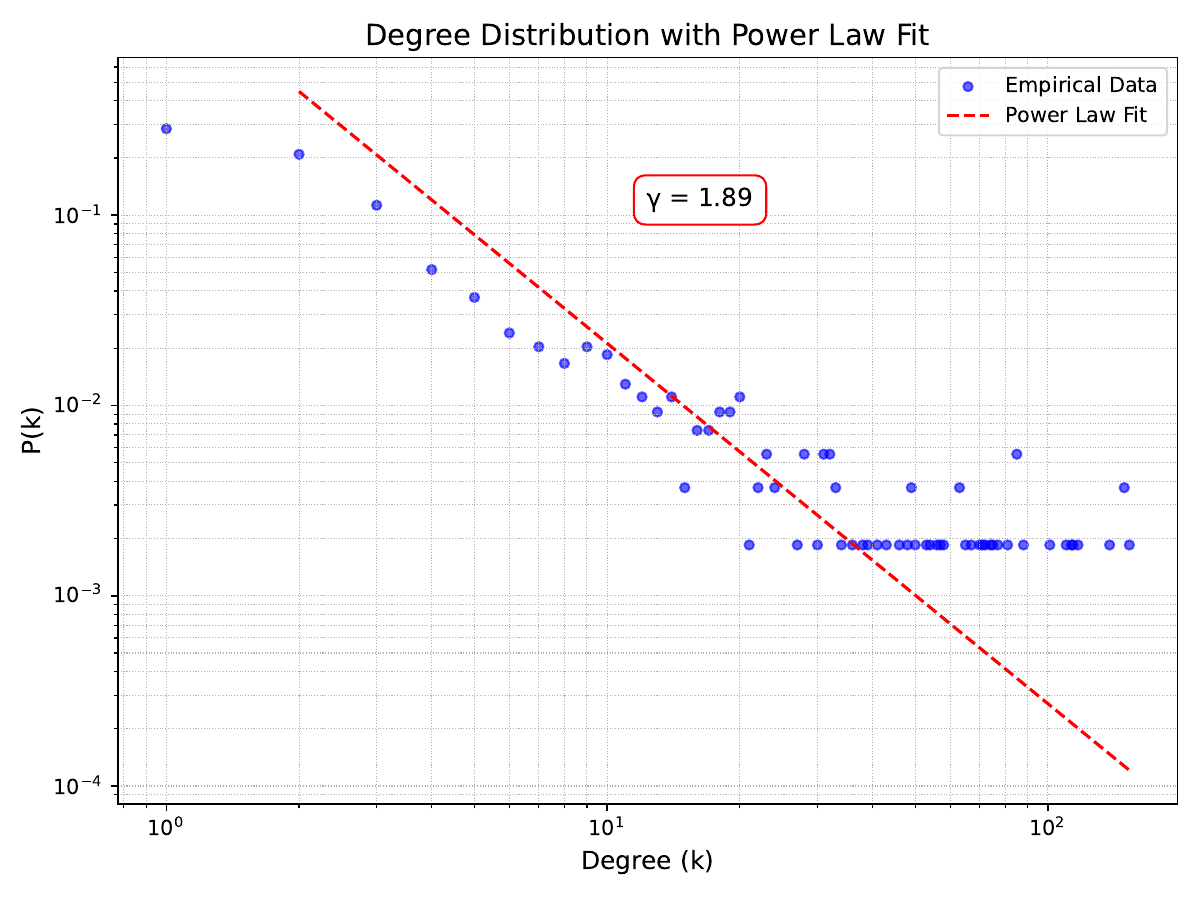}
    \caption{Degree Distribution with Power-Law Fit: \( \gamma \approx 1.89 \)}
    \label{fig:powerlaw}
\end{figure}
The degree distribution of USFN was analyzed on a log-log scale and fitted with a power-law model. As shown in Fig~\ref{fig:powerlaw}, the degree distribution closely aligns with power-law, with a fitted exponent \( \gamma \approx 1.89 \). The value of \( \gamma \) puts the USFN in a anomalous regime  \( \gamma \leq 2 \) . This regime is characterized with the biggest hub's degree growing linearly with the size of the network, i.e.,  \(k_{\text{max}} \sim N  \), with a hub and spoke configuration with average path independent to the size of the network \cite{barabasi2016network}.  These characteristics can be seen in Fig~\ref{fig:powerlaw} with a distinctive heavy-tail distribution where a small number of nodes consist of high degrees while the vast majority have only a few connections. 

However, there are tradeoff for these property, as networks dominated by high degree nodes are susceptible to targeted attacks, especially to the hubs, which will be further experimented in this paper. 

\subsection{Community Structure}
\begin{figure*}[htbp]
    \centering
    \includegraphics[width=0.7\textwidth, height=0.45\textheight]{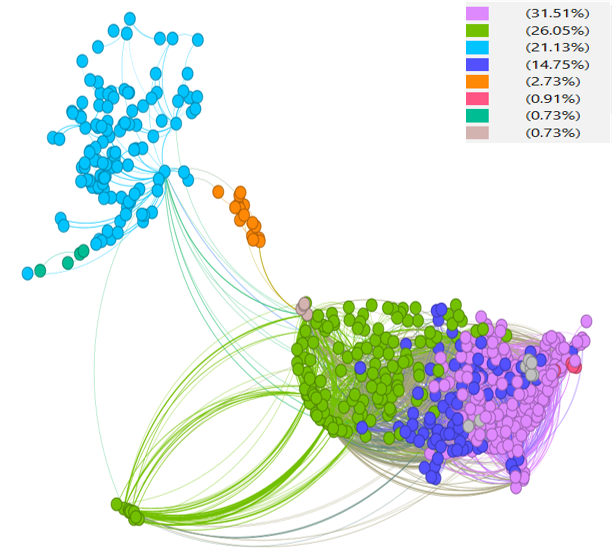}
    \caption{Communities structure with Louvain Algorithm}
    \label{fig:louvain}
\end{figure*}

Community structure in a network can be formulated using a clustering algorithm. The algorithm task is to assign a set of object to a community such that nodes in the communities are similar to each other than to those in other communities. 

In USFN we use two different type of clustering to create community structure and evaluate the communities.

\noindent 
\textbf{Louvain} algorithm, a hierarchical agglomerative methods that takes a greedy approach to local optimization is the first algorithm we use for creating communities\cite{emmons2016analysis}. It uses Newman-Grivan modularity in two steps. First it assigns each node to a community that will lead to an increase in modularity. It then create a super node from the communities created in the first step. The process repeats iteratively. 

\begin{table}[H]
\centering
\begin{tabular}{|c|c|c|c|}
\hline
\textbf{Community} & \textbf{Size} & \textbf{Edges} & \textbf{Density} \\
\hline
0 & 139 & 803 & 0.0837 \\
4 & 131 & 192 & 0.0225 \\
2 & 127 & 448 & 0.0560 \\
1 & 73  & 169 & 0.0643 \\
3 & 61  & 110 & 0.0601 \\
5 & 5   & 5   & 0.5000 \\
7 & 3   & 2   & 0.6667 \\
6 & 2   & 1   & 1.0000 \\
\hline
\end{tabular}
\caption{Community structure of USFN based on the Louvain clustering algorithm.}
\label{tab:louvain_communities}
\end{table}

The table~\ref{tab:louvain_communities} shows the number of communities and their respective size, edges and density which was then place into a geographically created network structure.

The community structure generated by the Louvain clustering algorithm reveals meaningful and geographically coherent groupings within the US Airport Network (USFN). As illustrated in Figure~\ref{fig:louvain}, nodes (airports) are clustered based on the density of connections, and the resulting communities largely align with physical regions of the United States.

To compare the community generated by the Louvain algorithm against some other algorithm, we use label propagation algorithm.

\noindent
\textbf{Label propagation algorithm} uses an iterative process to find stable communities. Each node in a graph start with a unique label and iteratively the nodes adopts the label of the most common among its neighbors.\cite{emmons2016analysis}

\begin{table}[H]
\small 
\centering
\caption{Comparison of Clustering Methods}
\begin{tabular}{|c|c|c|}
\hline
\textbf{Metric} & \textbf{Louvain} & \textbf{Label Propagation} \\
\hline
Modularity & 0.2818 & 0.0686 \\
\# Communities & 8 & 50 \\
Size Std & 66.3964 & 53.9321 \\
Edges Std & 330.0996 & 355.9516 \\
Density Std & 0.3371 & 0.2557 \\
\hline
\end{tabular}
\label{tab:cluster_comparison}
\end{table}

The table~\ref{tab:cluster_comparison} shows the difference between the communities formed by Louvain and label propagation. Label Propagation generates many more communities compare to the Louvain algorithm, with much less modularity suggesting that the communities created by label propagation doesn't exhibit many similar features.
\begin{figure*}[htbp]
    \centering
    \includegraphics[width=0.9\linewidth]{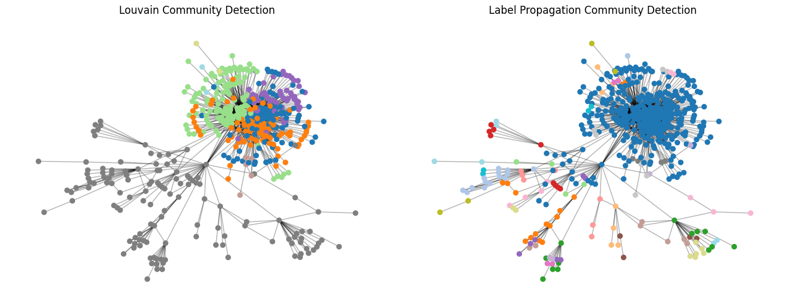}
    \caption{Communities Detected with Louvain Algorithm and Label Propagation Algorithm}
    \label{ComparisionL}
\end{figure*}

From the figure ~\ref{ComparisionL} we can see that there is a giant community that holds the majority of nodes in Label Propagation, with many small communities, which is not the case in Louvain, which has fewer communities with overall better well connected communities.

\begin{figure*}[htbp]
    \centering
    \includegraphics[width=0.9\linewidth]{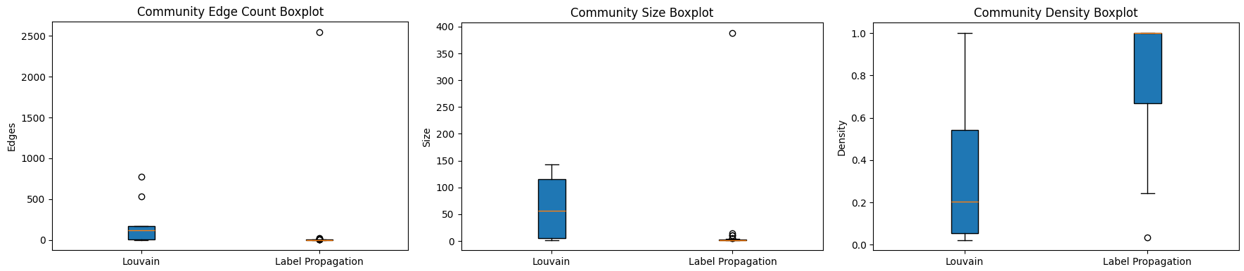}
    \caption{Comparison of community attributes using boxplots: Louvain vs. Label Propagation}
    \label{BoxPlotL}
\end{figure*}

From the box plot in figure ~\ref{BoxPlotL}, we see a box plot for edge, community size, and community density for Louvain and Label Propagation. 

Louvain shows a wider spread in edge counts across communities, while Label Propagation has very low and tightly clustered edge counts with one very large outlier. This shows Louvain was able to create a group with more interconnected nodes, while in label propagation, we have fragmented, loosely connected communities.

For community size, again, Louvain shows a broader distribution while Label propagation yields very small communities with many having very few nodes.

Label propagation produce many communities with almost 1 community density, suggesting all nodes are connected to each other, while it varies significantly in Louvain with a median around 0.2 and a long upper tail.
\subsection{Comparison with Null Models}
\begin{table*}[t]
\centering
\caption{Comparison of Real Network with Random ER and Scale-Free BA Models}
\begin{tabular}{|l|c|c|c|c|}
\hline
\textbf{Model} & \textbf{Nodes} & \textbf{Edges} & \textbf{Avg. Clustering (C)} & \textbf{Avg. Shortest Path (L)} \\
\hline
Real Network            & 541 & 2780 & 0.49 & 3.20 \\
\hline
Random ER Network 1     & 541 & 2767 & 0.02 & 2.95 \\
Random ER Network 2     & 541 & 2777 & 0.02 & 2.94 \\
Random ER Network 3     & 541 & 2832 & 0.02 & 2.92 \\
\hline
Scale-Free BA Network 1 & 541 & 2680 & 0.06 & 2.80 \\
Scale-Free BA Network 2 & 541 & 2680 & 0.06 & 2.80 \\
Scale-Free BA Network 3 & 541 & 2680 & 0.06 & 2.78 \\
\hline
\end{tabular}
\label{tab:combined_comparison}
\end{table*}

The structure of USFN was tested against two synthetic network: Random Erdős–Rényi (ER)\cite{erdos1959random}  and Scale-Free Barabási–Albert (BA)\cite{Baraba_si_1999} networks. The generated networks have the same number of nodes as the real network, while the edge count is a close approximation.
\subsubsection{Random Network Comparison}
We generated three Erdős–Rényi networks with 541 nodes and similar edge count as the real network (Table~\ref{tab:combined_comparison}) using probability: \[
p = \frac{2m}{N(N-1)}
\]
where, m is the number of edges and N is the number of nodes in the real network. 

\begin{figure}[H]
    \centering
    \includegraphics[width=0.9\linewidth]{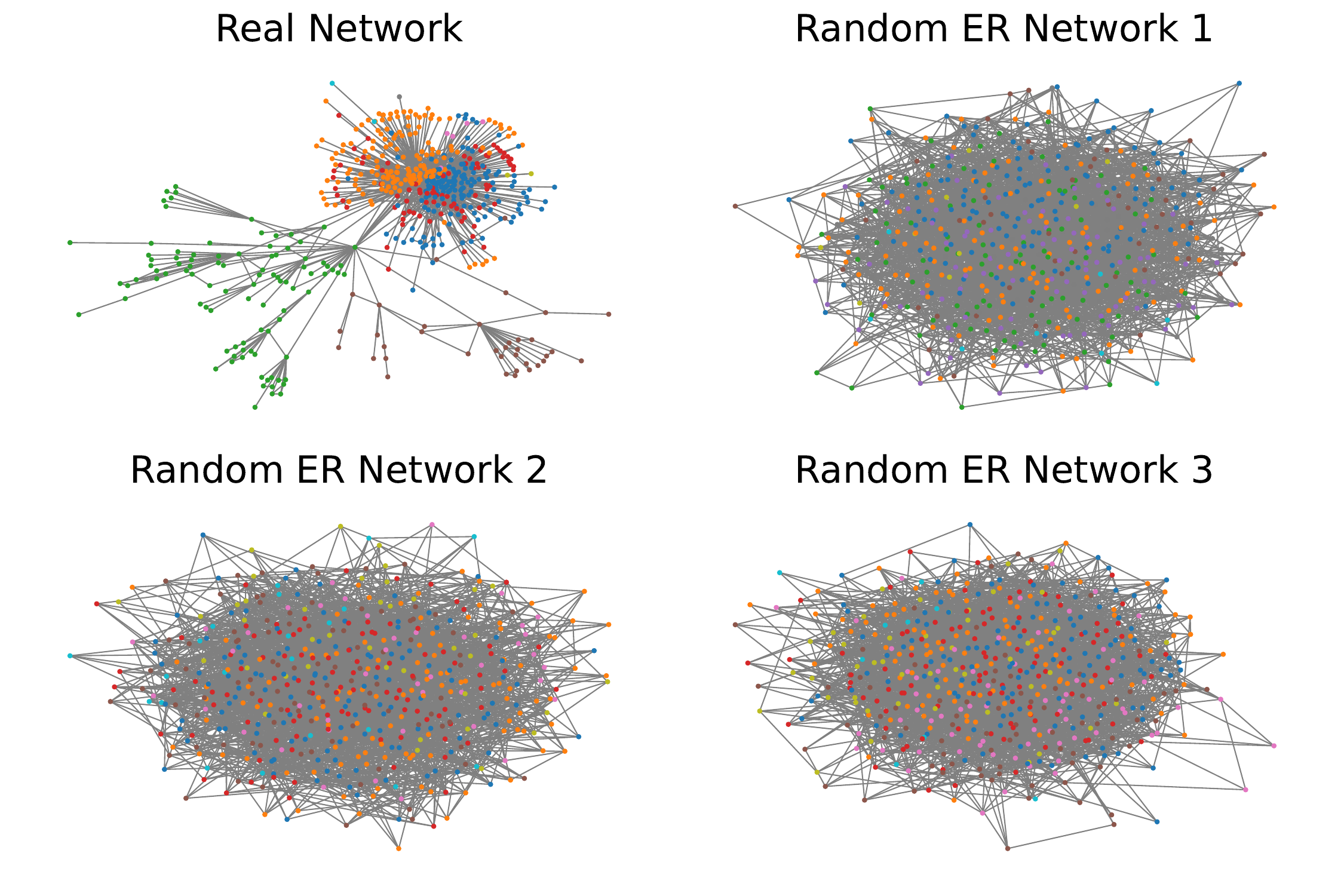}
    \caption{USFN vs Random Network}
    \label{USFNvsRandomNetwork}
\end{figure}

The results in Table~\ref{tab:combined_comparison} shows that the ER networks have shorter paths compared to USFN. This suggests that the USFN was probably not designed for the shortest distance but was more of an economical design. ER network also consistently have a low clustering coefficient (Fig~\ref{USFNvsRandomNetwork} ), compared to USFN's 0.49, suggesting the presence of communities that do not arise from random choice.
\begin{figure}[H]
    \centering
    \includegraphics[width=0.9\linewidth]{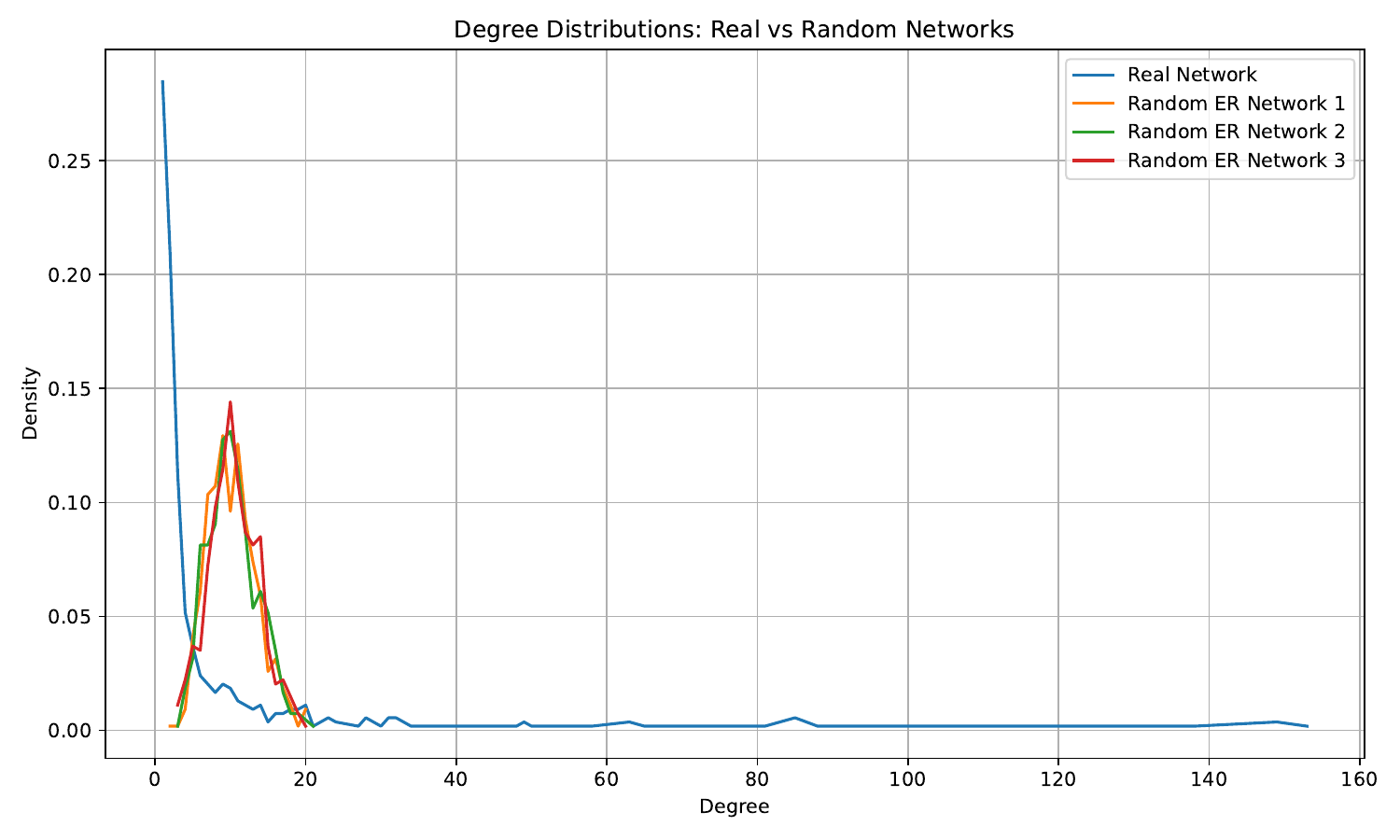}
    \caption{Degree Distribution: USFN Vs Random Networks}
    \label{RandomNetwork}
\end{figure}

The degree distribution for the generated random network compared with USFN network is given in Fig~\ref{RandomNetwork}. As expected, the random networks follows binomial distribution while the USFN network has a heavy-tailed distribution.  These results suggests that the USFN network is non-random.

\subsubsection{Scale-Free Network Comparison}
\begin{figure}[H]
    \centering
    \includegraphics[width=0.9\linewidth]{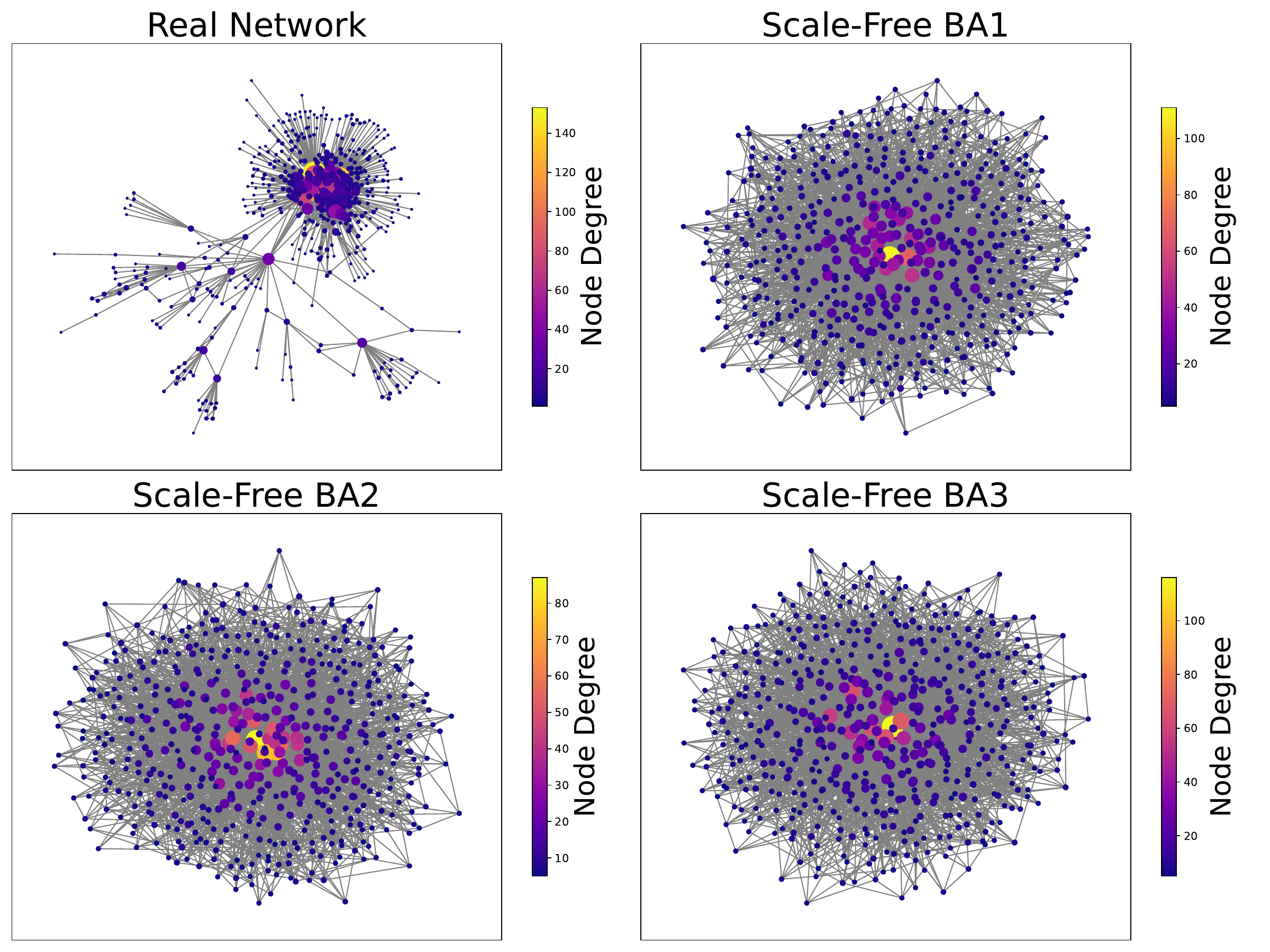}
    \caption{USFN vs Scale Free Network}
    \label{USFNvsScaleFree}
\end{figure}

Similarly, three scale free Barabási–Albert (BA) networks were generated with the same number of nodes (541). The generated networks here do not follow fixed edge probability but rather follows a preferential attachment mechanism\footnote{\href{https://networkx.org/documentation/stable/reference/generated/networkx.generators.random_graphs.barabasi_albert_graph.html}{Barabási–Albert Graph - NetworkX}}.
\begin{figure}[H]
    \centering
    \includegraphics[width=0.9\linewidth]{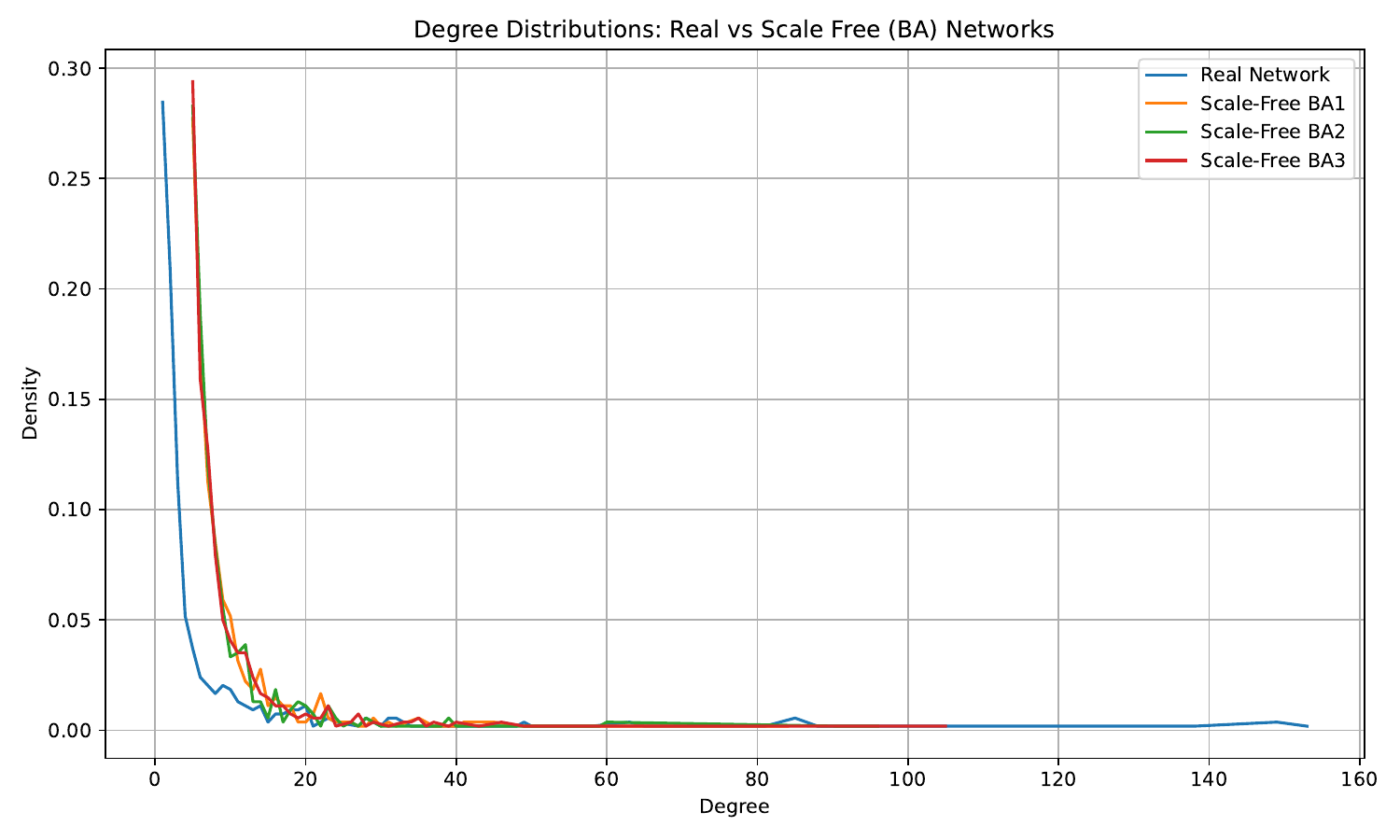}
    \caption{Degree Distribution: USFN Vs Scale-Free Networks}
    \label{Scale-Free}
\end{figure}
\vspace{-1em} 
These generated network replicates the heavy-tailed degree distribution nature of USFN (Fig~\ref{Scale-Free}) but with shorter path length. The BA networks also mimics the presence of hubs (illustrated Fig~\ref{USFNvsScaleFree}) like the USFN. However, the clustering coefficient was significantly lower (Table~\ref{tab:combined_comparison})  compared to USFN, suggesting that even with hubs in the BA model, it fall short in replicating the meaningful grouping of USFN.

\section{Percolation and Network Resilience}
For the purpose of determining how well the network holds up given random or targeted attack, nodes are removed, and we carry out various percolation and network resilience analyses. 

\subsection{Random Failure vs Targeted Attack}
We will use various centralities to determine the most significant nodes in our network and remove them so as to see if our network holds or collapse.

We will be looking at the change in the Largest Connected Component(LCC) and see changes in it as we remove a node. 

Largest Connected Component (LCC) in a network is the biggest subset of nodes where each node is reachable from any other node in that network.
\begin{table}[H]
\centering
\caption{Top 5 Nodes by Centrality Measures}
\label{tab:centrality_top5}
\begin{tabular}{|c|c|c|}
\hline
\textbf{Degree} & \textbf{Betweenness} & \textbf{Closeness} \\
\hline
ATL & ANC & DEN \\
ORD & DEN & ORD \\
DEN & ORD & LAS \\
DFW & SEA & MSP \\
MSP & ATL & SEA \\
\hline
\end{tabular}
\end{table}

From table~\ref{tab:centrality_top5} we determine the top 5 nodes for each of the centralities which we will use to see changes in Largest connected component in our network. 
\begin{figure*}[htbp]
    \centering
    \label{LCC}
    \includegraphics[width=0.9\linewidth, height=0.45\textheight]{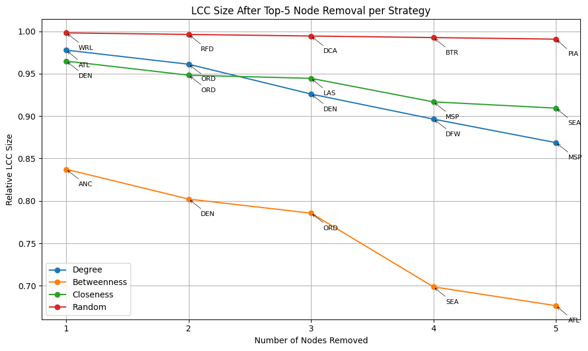}
    \caption{LCC Size change after node removal}
    \label{0.15}
\end{figure*}

From the chart above we can see that removing nodes with highest betweenness centrality has the most destructive impact. LCC size has a sharp drop with just removing 5 nodes. Removing nodes with highest degree centrality and closeness centrality had similar impact with gentle decline in LCC. Removing random node had barely any effect after removing 5 nodes.

This suggest that our network while is resistant to random attack can collapse with targeted attack with highest damage when attacked by removing nodes with highest betweenness centrality.
\subsection{Resistance to Cascade of Failure Propagation}
\begin{figure*}[htbp]
    \centering
    \includegraphics[width=0.9\linewidth, height=0.45\textheight]{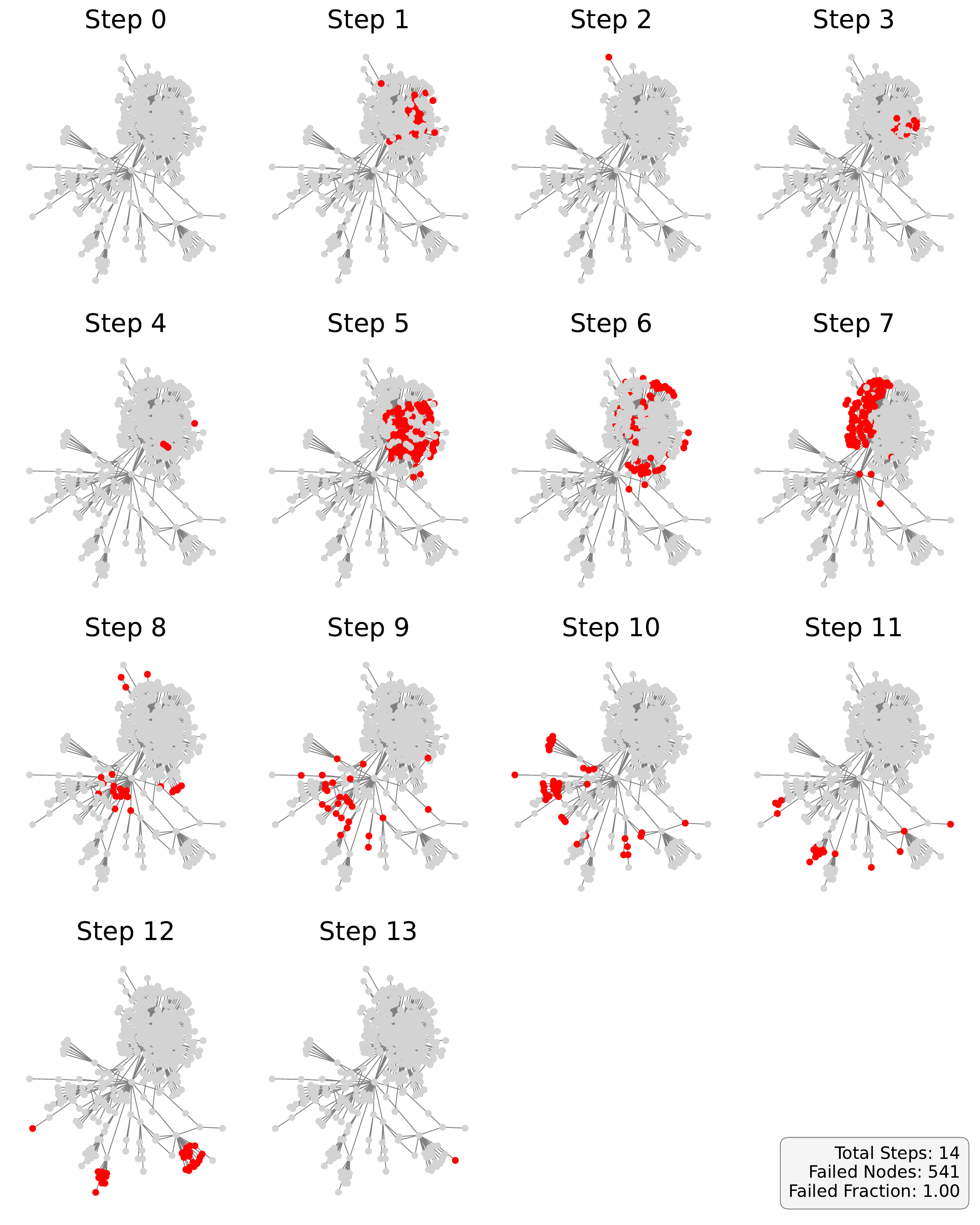}
    \caption{Cascade failure for threshold \( = 0.15 \) \(\left(\frac{3}{20}\right)\)}
    \label{0.15}
\end{figure*}
We stimulated cascading failures in the USFN to test its robustness. The node with highest-degree (hub) was chosen as the origin of  failure to stimulate a worst-case scenario. For each simulation, a node fails if it loses a fraction of its neighbors exceeding the breakdown threshold. The breakdown threshold was tested for a range from 0.05 to 0.50, and the average fraction of failed nodes was recorded. 
\begin{figure}[H]
    \centering
    \includegraphics[width=0.9\linewidth]{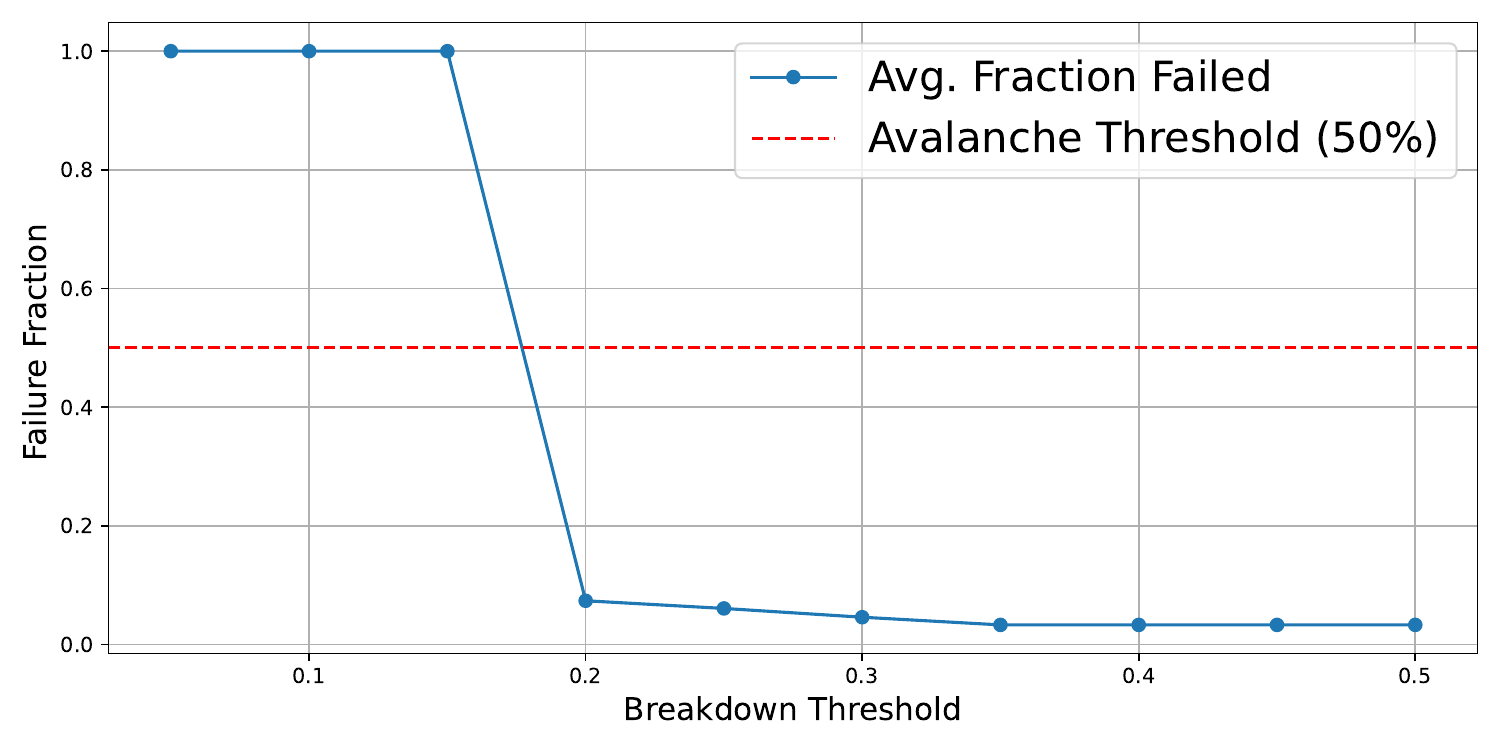}
    \caption{Critical Breakdown Threshold}
    \label{BreakdownThreshold}
\end{figure}
As shown in Fig~\ref{BreakdownThreshold}, there is a sharp transition in the network vulnerability. For threshold \( \leq  \) 0.15, we see that 100\% (541) of the nodes fail in just 14 steps. This shows that the failure is propagated throughout the USFN (as shown in Fig~\ref{0.15}), forming a complete cascade failure.

\begin{figure}[H]
    \centering
    \includegraphics[width=0.9\linewidth]{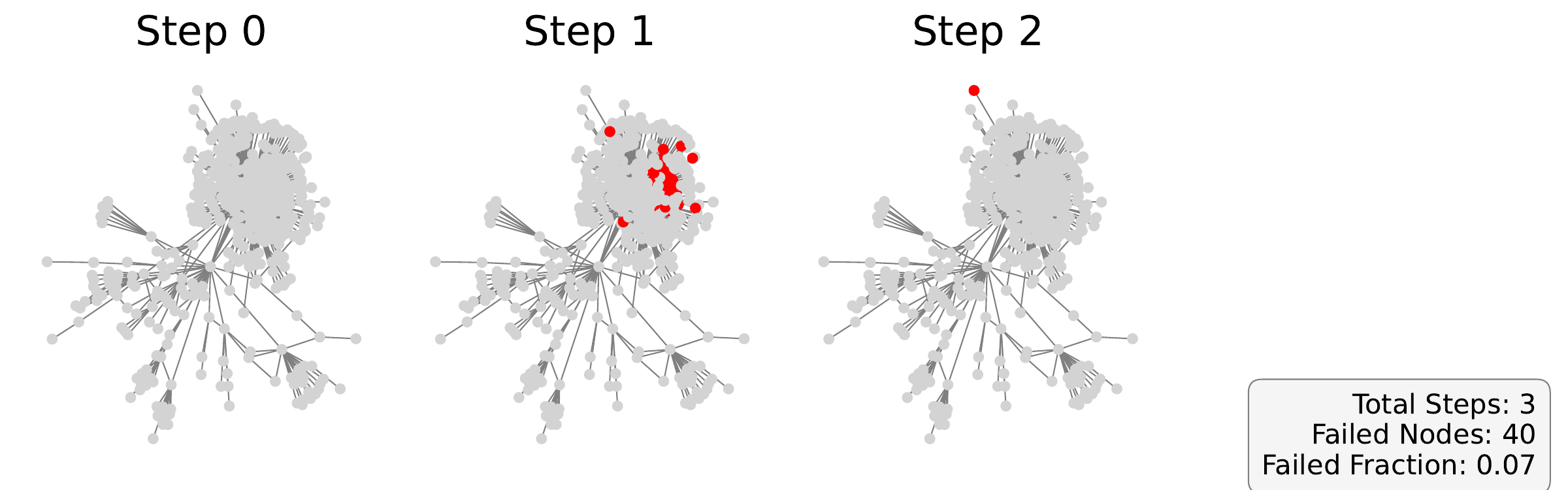}
    \caption{Cascade failure for threshold \( = 0.20 \) \(\left(\frac{1}{5}\right)\)}
    \label{0.20}
\end{figure}

However, for threshold  \( \geq  \) 0.20, the fraction of failed nodes drop to 7\% with failure contained in just 3 steps and only 40 nodes affected in total (Fig~\ref{0.20}). Thus, the network shows strong resilience for this threshold, and with only around 6\% nodes affected for thresholds \( \geq  \) 0.25, with no simulations resulting in large scale failure. 
These results show that the critical breakdown threshold for the USFN lies between 0.15 and 0.20. USFN is resilient for threshold values greater than the critical range, but is prone to cascading failure below the range. As the critical threshold is on the lower side, it implies high fragility such that if one major hub (e.g., ATL, ORD, DFW) fails, it can quickly trigger cascading shutdowns, which in turn can disrupt other hubs and eventually the entire USFN.

\section{Discussion}
The USFN shows scale-free properties with degree distribution closely following power-law but falls under the anomalous regime \( \gamma \approx 1.89 \). This shows the presence of few highly connected hubs with many other low-degree nodes, which makes for an efficient routing system comparable to the Australian Airport Network with \( \gamma \approx 1.12 \) \cite{HOSSAIN20171}.

Compared to null models, the USFN stands out due to it's meaningful community structure (clusters). The Erdős–Rényi random networks lacked clustering and hub formation, while Barabási–Albert scale-free models replicated the heavy-tailed distribution, it still had a low clustering coefficient. This highlighted that the airport network is constrained by geography, policies, or other real-world constraints and is not random.

The Louvain algorithm generated more interpretable and heterogeneous community structures within the U.S. airport network. It had a broader distribution of edge count and size per community, suggesting that the airport network communities are within structural and functional constraints. This was not the case with Label propagation, which produce many small, dense, and uniform clusters with very few large outliers.

Having scale free network like nature USFN is more vulnerable to targeted attack specially if we remove nodes based on the highest betweenness centrality. This indicate that these nodes serve as critical bridges between various region. The comparison on ~\ref{LCC} highlights that the airport network is vulnerable to targeted failures while being robust to random attack.

Cascading failure simulations revealed that the network is fragile given the presence of hubs. For a threshold below 0.15, failure of just one major hub can create a total collapse, while a threshold greater than 0.20 maintains the network's stability. This highlights the resilience boundary of USFN and the risk posed by critical node failures.

\section{Conclusion}
This study provides a comprehensive analysis of U.S. Flight Network (USFN) using complex network science. The results show that USFN is an efficient system due to the presence of meaningful community structures (clusters), yet is a vulnerable system due to it's scale-free nature, making it susceptible to targeted attacks.

\section{Future Work}
Building on top of what we already have we can see the changes in the resilience evolve over time and see if it has changed or remain the same. With a weighted edge we can have more insight into the traffic distribution and importance of the node. Further more we can develop optimization strategies like a new route or decentralizing traffic can be done. Likewise, we can integrate the air network with other transportation modes like rail and road, which could provide insight into multi modal infrastructure system.

\bibliographystyle{plain}
\bibliography{ref}

\end{multicols}
\end{document}